\documentclass[twocolumn,showpacs,amsmath, amssymb, aps, prl]{revtex4-1}
\usepackage{graphicx}
\usepackage{dcolumn}
\usepackage{bm}

\begin{document}
\title{Casimir electromotive force in periodic configurations}
\author{Evgeny\,G.\,Fateev}
 \email{e.g.fateev@gmail.com}
\affiliation{%
Institute of mechanics, Ural Branch of the RAS, Izhevsk 426067, Russia
}%
\date{\today}
\begin{abstract}
The possibility in principle of the existence of Casimir electromotive force 
(EMF) is shown for nonparallel nanosized metal plates arranged in the form 
of a periodic structure. It is found that EMF does not appear in strictly 
periodic structures with parallel plates. However, when the strict 
periodicity is disturbed in nonparallel plates, EMF is generated, and its 
value is equal to the number of pairs of plates in a configuration. 
Moreover, there are some effective parameters of the configuration (angles 
between plates, plate lengths and length to length ratios), at which the EMF 
generation per unit of the length of the periodic structure is maximal.
\end{abstract}

\maketitle
\textbf{INTRODUCTION}

Recently in work \cite{Fateev:2015}, the possibility in principle of the 
existence of Casimir electromotive force has been shown for a single open 
perfectly conducting nanosized structure with nonparallel plates (wings). 
Naturally, in parallel metal plates in the classical configuration studied 
by Casimir \cite{Casimir:1948, Casimir:1949, Milton:2001, 
Klimchitskaya:2009, Bordag:2009}, electromotive force must 
not generate. However, at the ends of the plates, there can be certain 
fluctuations of electric potentials due to Johnson-Nyquist thermal noise 
\cite{Bimonte:2008} and interference currents because of radio 
interference.

The possibility of Casimir EMF generation is associated with an effect 
similar to light-induced electron drag, which can appear in metals 
\cite{Gurevich:1992, Shalaev:1992, Shalaev:1996}, 
graphite nano-films \cite{Mikheev:2012} and semiconductors 
\cite{perovich:1981}. In our case, the drag effect can arise in 
nonparallel and perfectly conducting wings due to the total uncompensated 
action of virtual photons upon electrons. Earlier it has been shown that in 
addition to the EMF generation in its wings, the system with nonparallel 
wings can also experience Casimir expulsion force 
\cite{Fateev:2012, Fateev:2013} and other interesting effects 
\cite{Fateev:2014}. The force shows up as a time-constant expulsion of 
nonparallel plates in the direction of the smallest angle between them. It 
should be noted that this force significantly differs from Casimir pressure 
and expulsion which are capable of creating only effects of levitation above 
body-partners \cite{Jaffe:2005, Leonhardt:2007, 
Levin:2010, Rahi:2010, Rahi:2011}. The uncompensated 
action of the forces upon nonparallel structures is due to the nonuniform 
action of Casimir forces upon the opposite ends of the configuration 
asymmetrical along one of the coordinates. Optimal parameters have been 
found for the opening of the angles between the plates and for the plate 
lengths, at which expulsion forces and EMF obtained should be maximal. 

\begin{figure}
\hypertarget{fig1}
\centerline{\includegraphics[width=2in,height=2.2in]{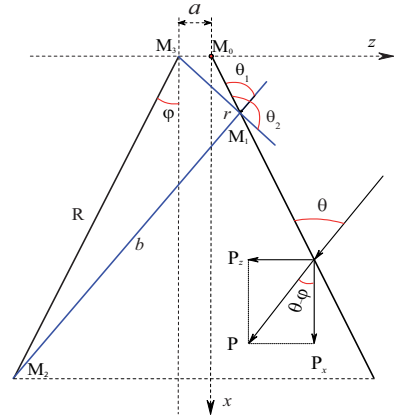}}
\caption{Schematic view of the configuration with nonparallel metal plates with the 
surface length $R$, the particular cases of which are parallel plates for 
$\varphi =0$ and a triangle at $a=0$. The configuration with the width $L$ in 
the $y$-direction (perpendicular to the figure plane) is shown in the 
Cartesian coordinates in the plane $(x,z)$. The blue lines designate the 
virtual rays with the length $b$-outgoing from the point $\mbox{M}_1 $ under 
the limit angles $\Theta _1 $ and $\Theta _2 $ toward the right plate and 
finishing at the ends of the opposite wing at the points $\mbox{M}_2 $ and 
$\mbox{M}_3 $, respectively.}
\end{figure}
\begin{figure*}
\hypertarget{fig2}
\centerline{\includegraphics[width=4in,height=2.2in]{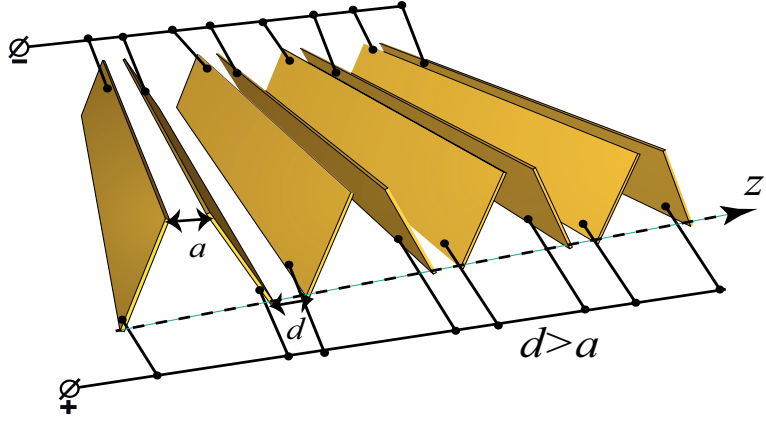}}
\caption{Schematic view of the periodic configuration with pairs of 
nonparallel wings in the $z$-direction at the distance $d$ from one another. }
\end{figure*}
For obtaining large total uncompensated forces and EMF, it is important that 
the conditions for the existence of the discussed effects in periodic 
structures are investigated. In Ref. \cite{Casimir:1949}, for example, 
it is shown that there are no uncompensated forces in strictly periodic 
structures with parallel plates which are equally spaced. The expulsion 
effect arises when the wings are unparallel and the separation distance 
between the periods is larger than the minimal distance $a$ between the 
wings. There is an optimum of the $d/a$ relation at which the maximal effect 
is obtained.

It is natural to ask if the EMF excitation is possible in periodic 
structures constructed on the basis of configurations having the effect of 
Casimir EMF.

\textbf{THEORY}

Let us consider a periodic configuration with nonparallel nanosized metal 
plates for studying the possibility of the Casimir EMF existence. The inner 
and outer surfaces of the plates should have the properties of mirrors with 
the reflection coefficient $\rho $. The configuration can completely be 
merged into a material medium or be its part with the parameters of 
dielectric permeability different from those of physical vacuum. In a 
Cartesian coordinate system, a single nonparallel configuration looks like 
two thin metal plates with the surface width $L = 1$ m (oriented along the 
$z$-axis) and length $R$ at the distance $a$ from one another; the angle $\varphi$ 
between them can be changed (at the same time and by the same value for both 
wings) as it is shown in \hyperlink{fig1}{Fig.1.}

In the Cartesian coordinates, the periodic configuration with pairs of 
nonparallel wings looks like it is shown in \hyperlink{fig2}{Fig.2.} Each figure in the 
configuration period is completely similar to the single figure depicted in 
\hyperlink{fig1}{Fig.1.} In the period, between the ends of the figures there is the distance $d$.

In the first approximation, Casimir EMF for one nonparallel wing 
$\Delta E_\parallel$ can be 
found in the form \cite{Fateev:2015}
\begin{equation}
\label{eq1}
\Delta E_\parallel =\frac{1}{2n_b e}\left[ {1-\rho -k} \right]\int_0^L {dy} 
\int_0^{r_{\max } } {P(\Theta ,r,\phi )dr} .
\end{equation}
Here $n_b $ is volume density, $e$ - electron charge, $\rho $ - reflection 
coefficient, $k$ - photon transmission, $P(\Theta ,r,\phi )$- local specific 
pressure at each point $r$ on the wing with the length $r_{\max } $ and 
width $L$
\begin{equation}
\label{eq2}
\begin{array}{c}
P(\Theta ,r,\phi )=\frac{\hbar c\pi ^2}{240 s^4}\int_{\Theta _1 }^{\Theta 
_2 } {d\Theta } \sin (\Theta -2\varphi )^4\sin \Theta \cos \Theta \\ 
 =-\frac{\hbar c\pi ^2}{240 s^4}A(\varphi ,\Theta _1 ,\Theta _2 ), \\ 
 \end{array}
\end{equation}
where
\begin{equation}
\label{eq2}
\begin{array}{c}
 A(\varphi ,\Theta _1 ,\Theta _2 )=\frac{1}{96}\left[ {24\Theta _1 \sin 4\varphi 
-24\Theta _2 \sin 4\varphi } \right. \\ 
 +18\cos 2\Theta _1 -18\cos 2\Theta _2 \\ 
 +6\cos (4\varphi -4\Theta _2 )-6\cos (4\varphi -4\Theta _1 ) \\ 
 +3\cos (8\varphi -2\Theta _2 )-3\cos (8\varphi -2\Theta _1 ) \\ 
 \left. {+\cos (8\varphi -6\Theta _1 )-\cos (8\varphi -6\Theta _2 )} \right]. \\ 
 \end{array}
\end{equation}
\begin{figure*}
\hypertarget{fig3}
\centerline{
\includegraphics[width=1.6in,height=1.6in]{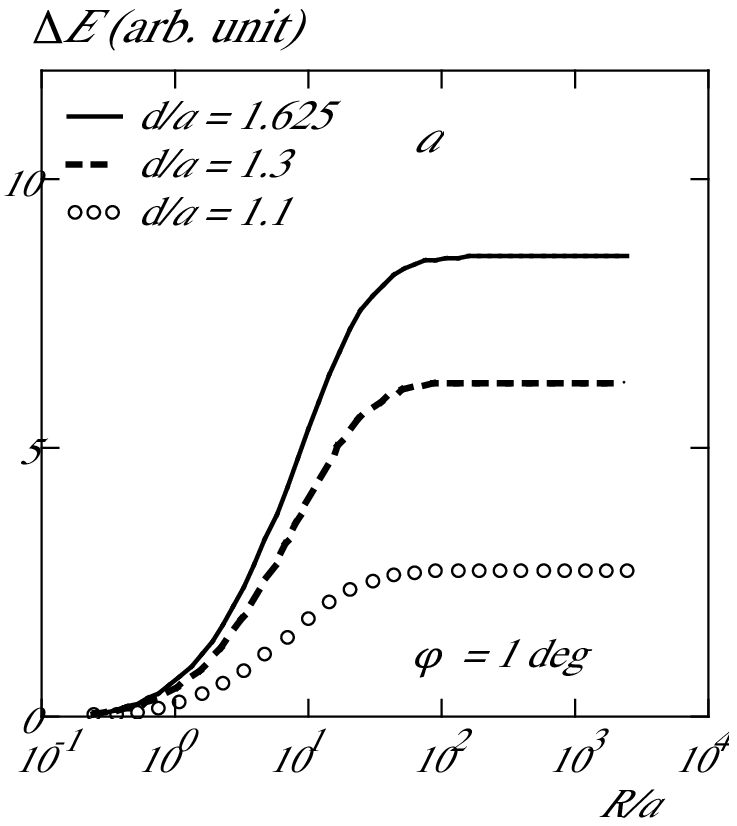}
\includegraphics[width=1.6in,height=1.6in]{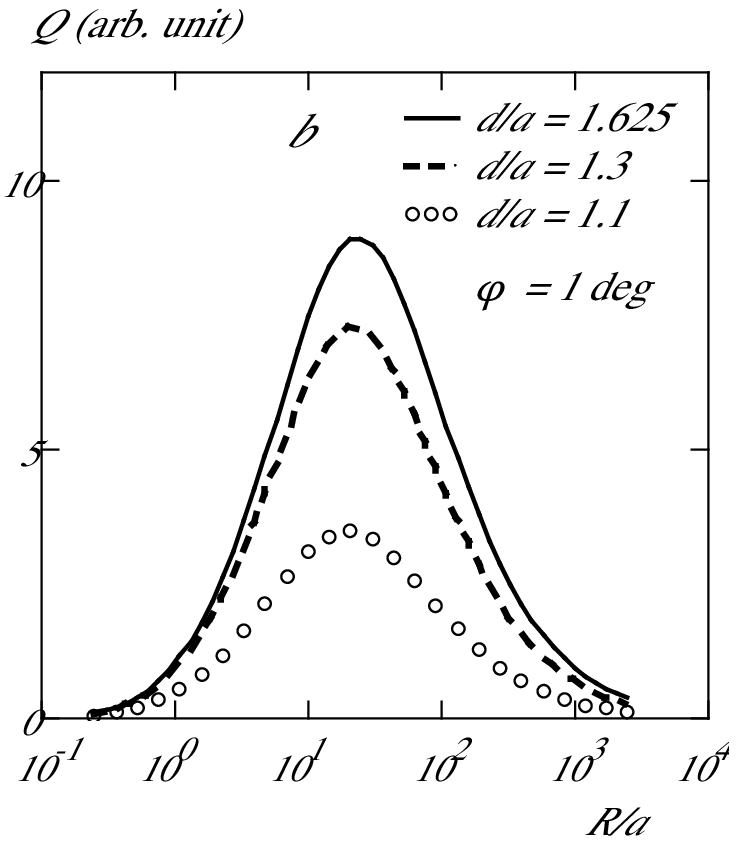}
\includegraphics[width=1.6in,height=1.6in]{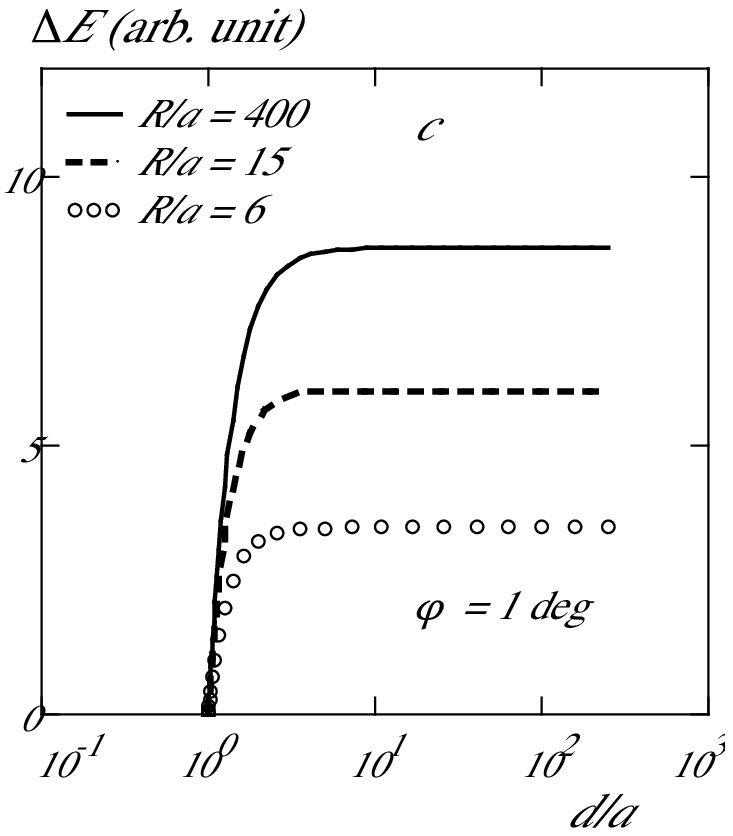}
\includegraphics[width=1.6in,height=1.6in]{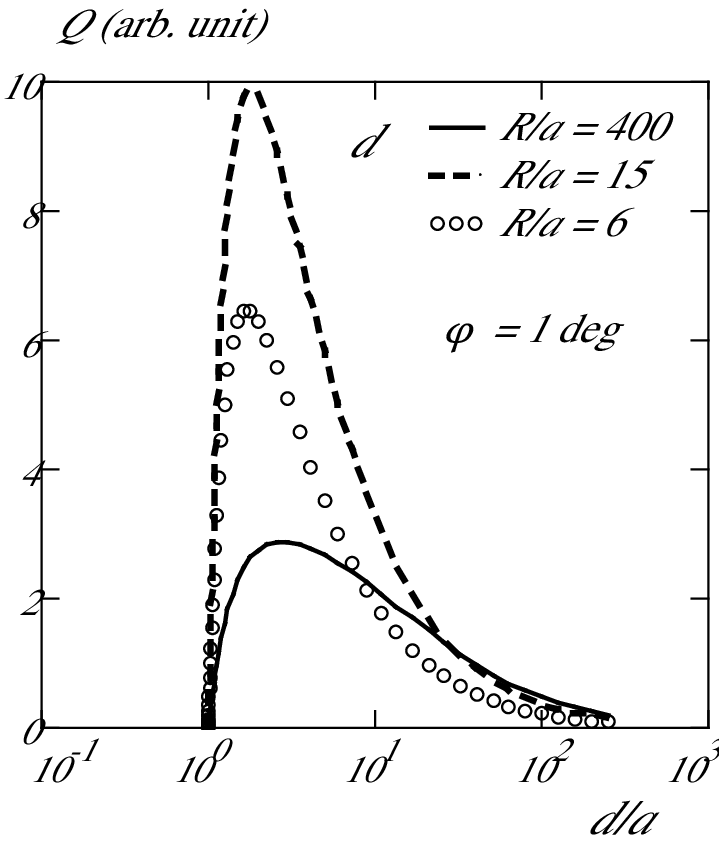}}
\label{fig3}
\caption{The total EMF in the periodic configuration at series connection of the 
wings in the chain depending on the relation $R/a$ in the structure ($a)$ and 
the effectiveness $Q$ of the structure ($b)$ at different relations of $d/a$ at 
$\varphi =1\;\deg $}
\end{figure*}
\begin{figure*}
\hypertarget{fig4}
\centerline{
\includegraphics[width=1.6in,height=1.6in]{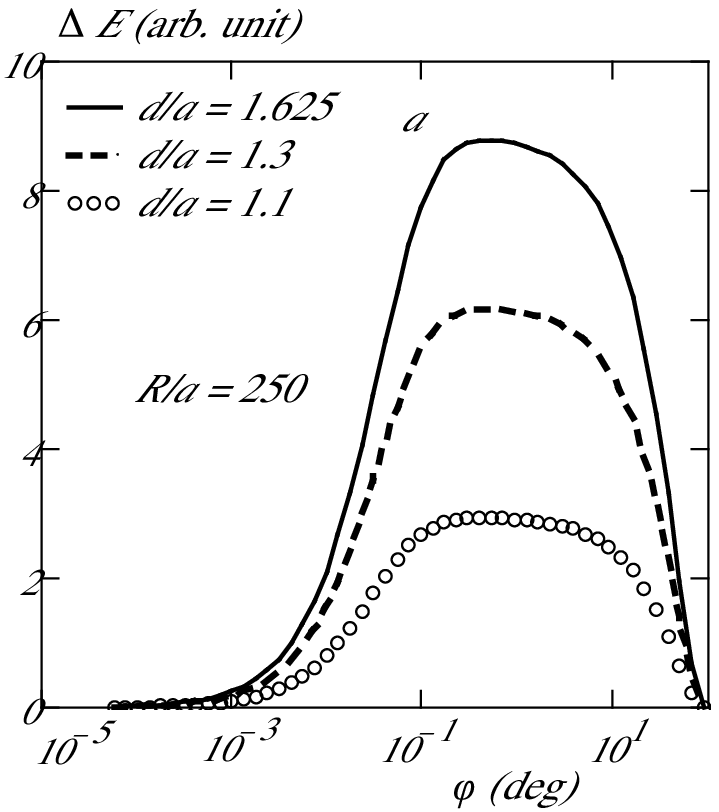}
\includegraphics[width=1.6in,height=1.6in]{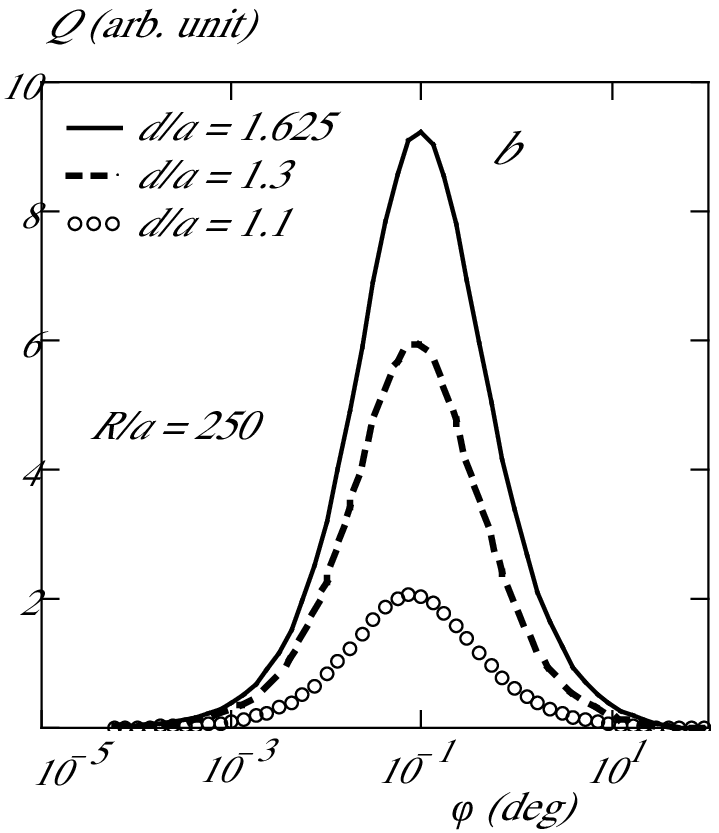}
\includegraphics[width=1.6in,height=1.6in]{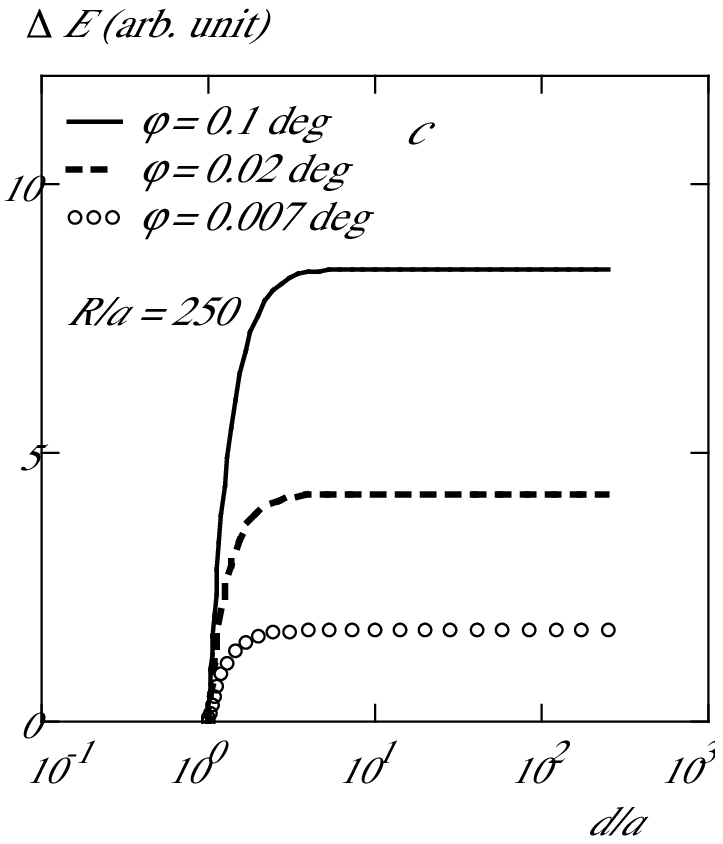}
\includegraphics[width=1.6in,height=1.6in]{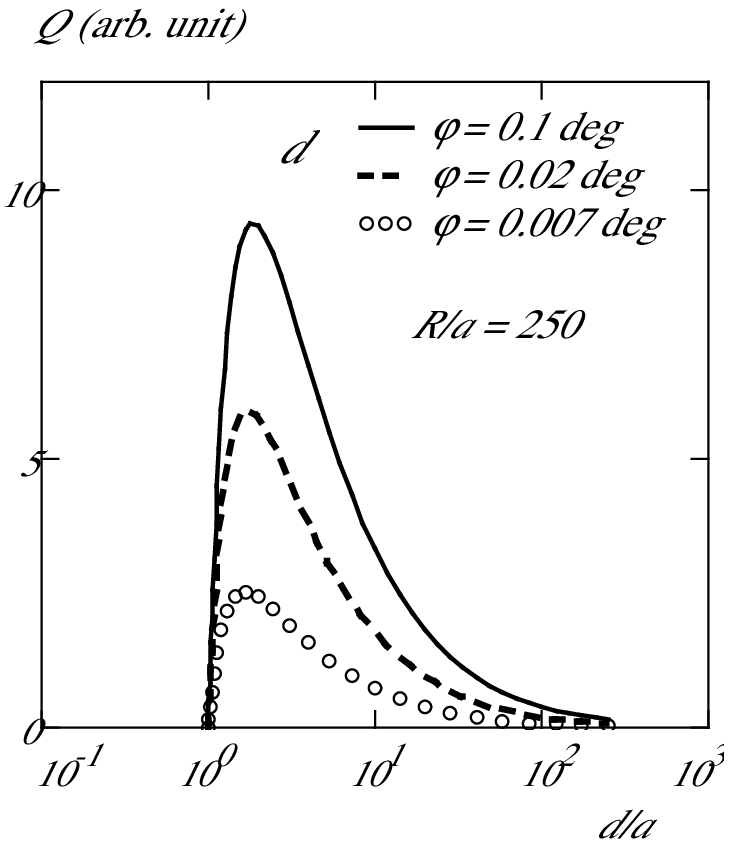}}
\label{fig4}
\caption{The total EMF in the periodic configuration at series connection in 
the chain of the wings depending on the angle $\varphi$ in the structure ($a)$, 
and the structure effectiveness $Q(b)$ at different relations $d/a$ and 
$R/a=250$. The total EMF ($c)$ for different angles $\varphi $ depending on the 
relation $d/a$ at $R/a=250$.}
\end{figure*}
\begin{figure*}
\hypertarget{fig5}
\centerline{
\includegraphics[width=1.6in,height=1.6in]{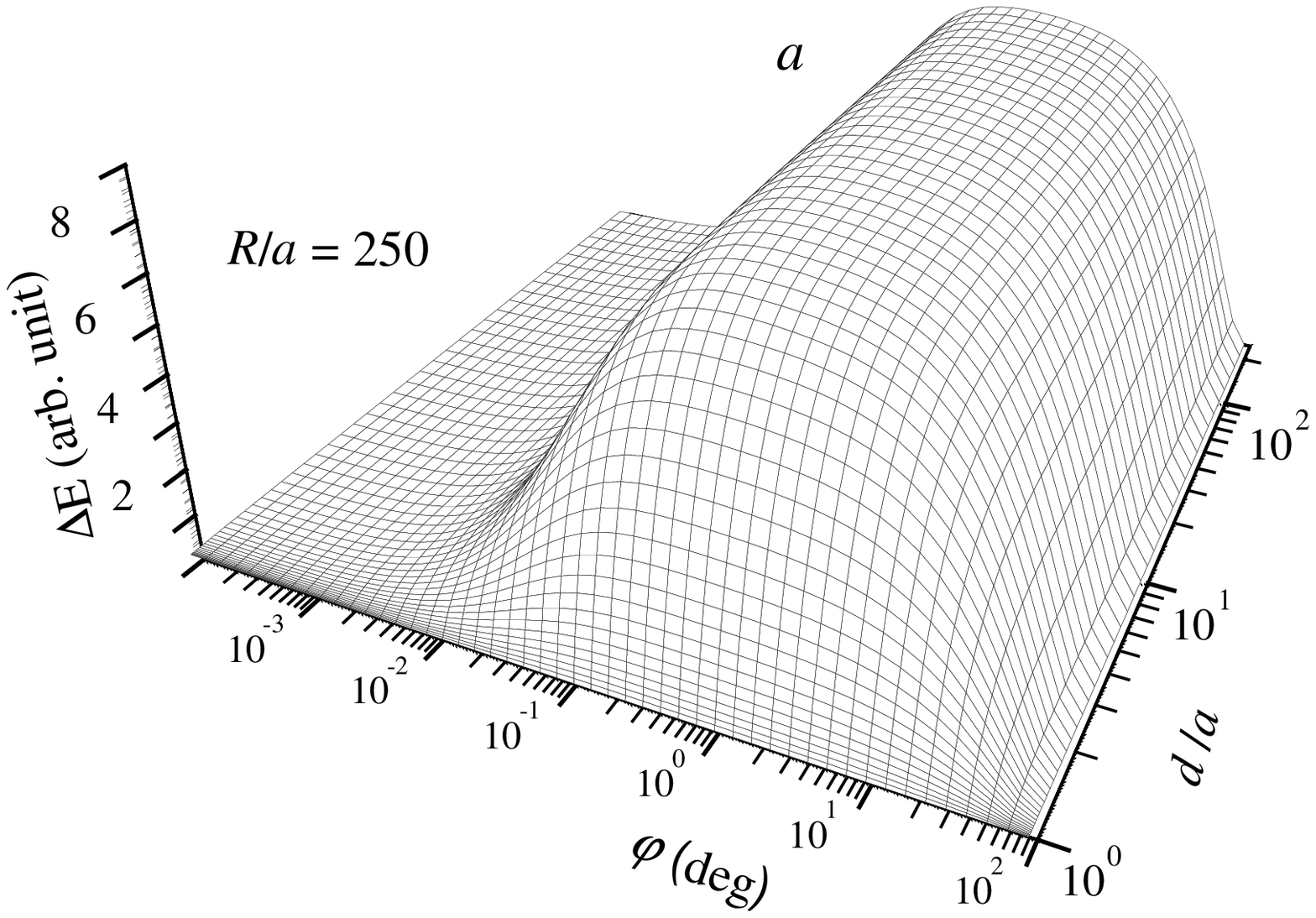}
\includegraphics[width=1.6in,height=1.6in]{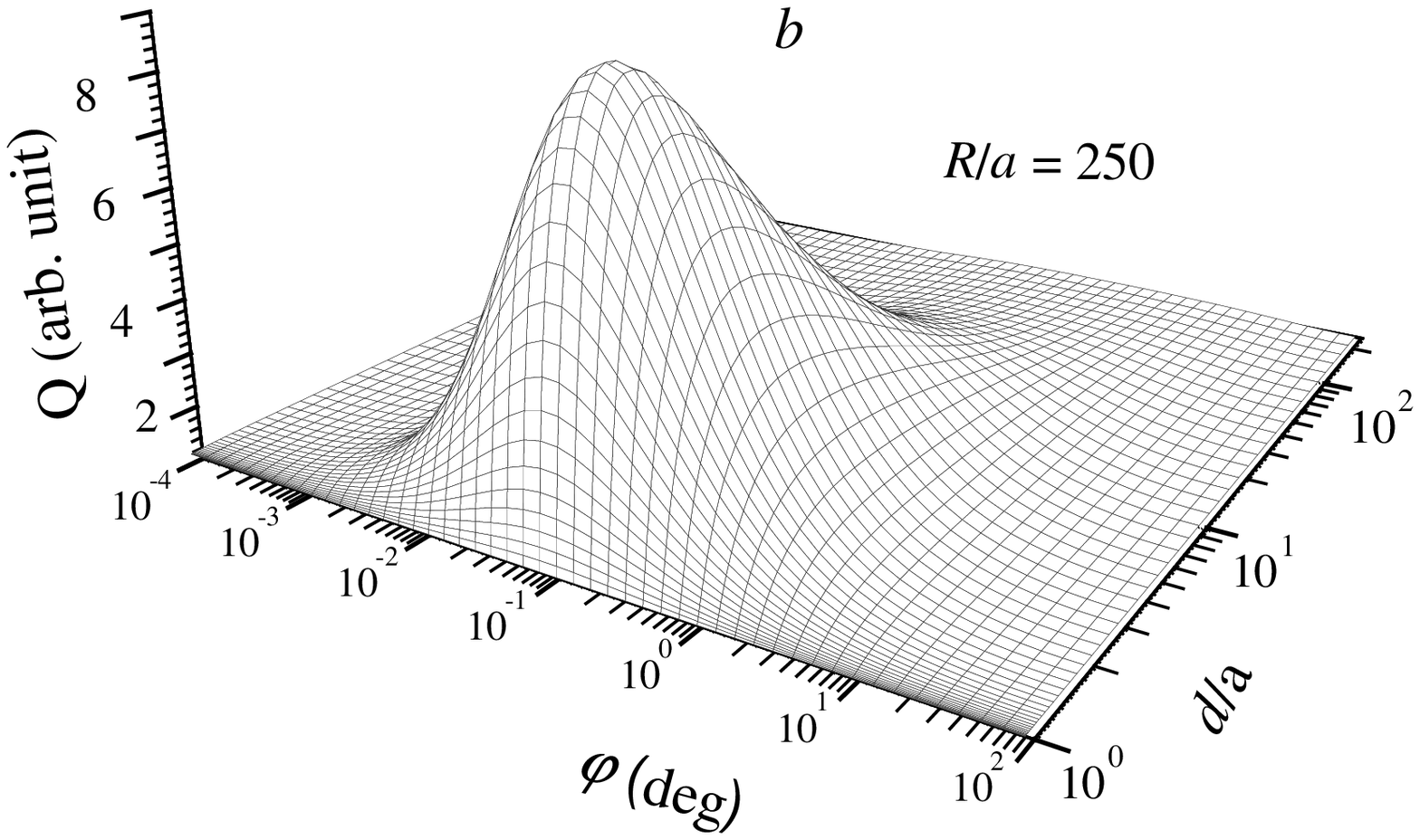}
\includegraphics[width=1.6in,height=1.6in]{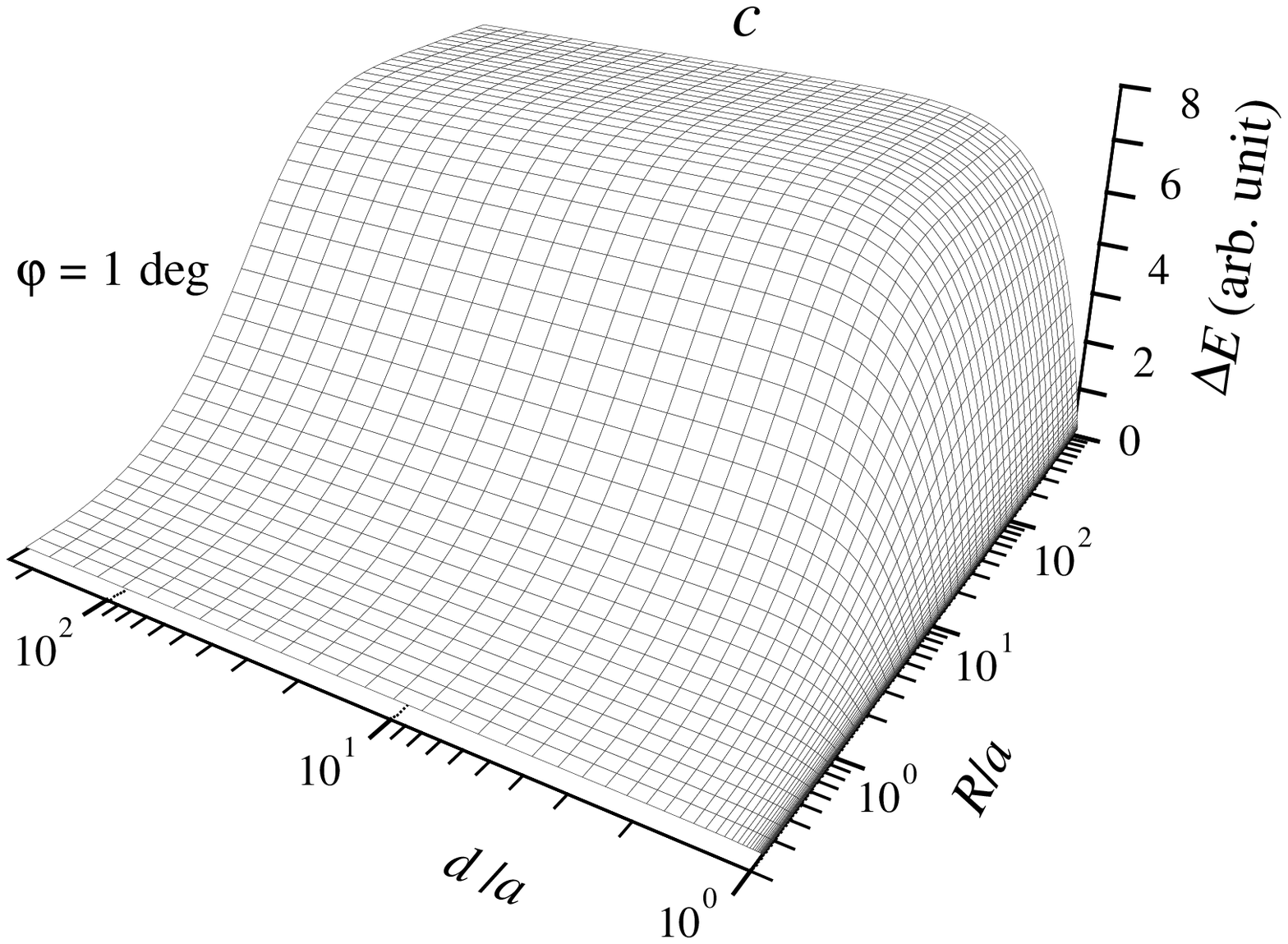}
\includegraphics[width=1.6in,height=1.6in]{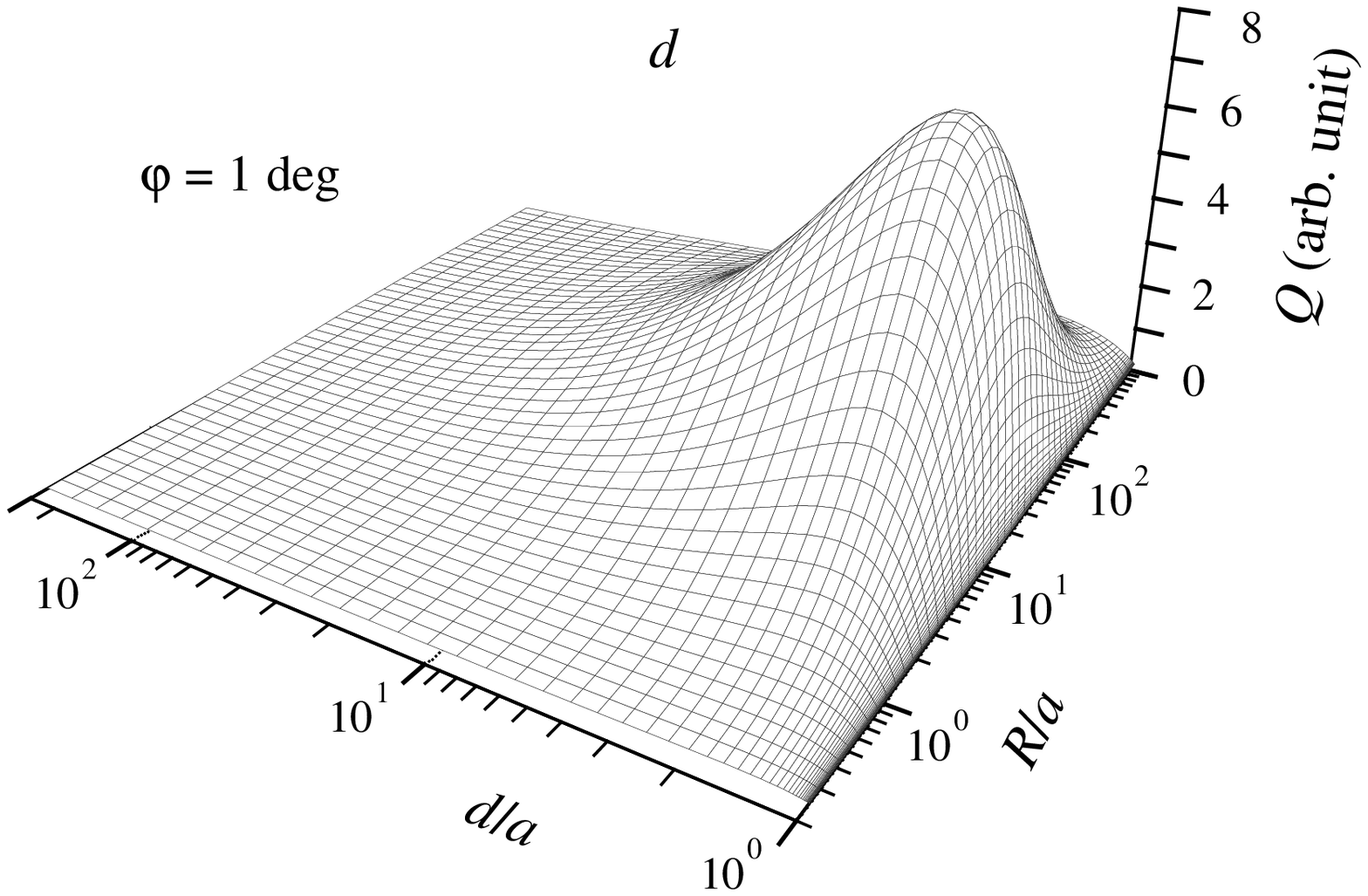}}
\label{fig5}
\caption{The total EMF in the periodic structure at series connection in the 
chain of the wings depending on the angle $\varphi $ and the relation $R/a$ in 
the structure ($a)$, and the corresponding structure effectiveness $Q$ ($b)$. The 
total EMF ($c)$ for different angles $\varphi $ depending on the relations $d/a$ 
and $R/a$, and the corresponding structure effectiveness $Q$.}
\end{figure*}
In formula (2), $\hbar =h/2\pi.$ is reduced Planck constant, $c$ -- speed of light. The 
functional expressions for limit angles between the wings $\Theta _1 
,\;\Theta _2 $ and the $s$ parameter have the following 
form
\begin{equation}
\label{eq4}
\begin{array}{c}
\Theta _1 =\mbox{arccos}\left[ {-\frac{r+a\sin \varphi -R\cos 2\varphi }{\sqrt 
{\left( {a+R\sin \varphi +r\sin \varphi } \right)^2+\left( {r\cos \varphi -R\cos \varphi 
} \right)^2} }} \right],
 \end{array}
\end{equation}
\begin{equation}
\label{eq5}
\Theta _2 =\mbox{arccos}\left[ {-\frac{r+a\sin \varphi }{\sqrt {a^2+r^2+2ra\sin 
\varphi } }} \right],
\end{equation}
\begin{equation}
\label{eq6}
s=\frac{\sin (2\varphi -\Theta _2 )(a+r\sin \varphi )}{\sin (\varphi -\Theta _2 )}^.
\end{equation}

Thus, here the scheme is presented for calculating the EMF generation due to 
virtual photons in nanosized metal configurations in optical approximation.
When such configurations are arranged in a periodic structure, it is 
necessary to take into consideration the following. The periodic arrangement 
of $s$ pairs of nonparallel wings with distance $d$ between them and against 
the $x$-axis with their wider sides leads to the formation of $n-1$ pairs of 
wings with oppositely directed sides (see \hyperlink{fig2}{Fig.2}). In this case one wing of 
each configuration is at the same time a wing of the other configuration the 
wide opening of which is oppositely directed. Thus, for $n$ pairs of 
configurations periodically arranged along the $z$-axis it is possible to 
write the following expression for their total EMF \cite{Rizzoni:2008} 
(if the wings as the sources of EMF are connected in a chain in series and 
not in parallel)
\begin{equation}
\label{eq7}
\Delta E_{total} =\sum\limits_{n=1}^n {\Delta {E_n} } ={n\Delta E(a)}-{(n-1) \Delta E(d)}
\end{equation}
Here $\Delta E(a)$ is the EMF for the shortest distance $a$ between the pair 
of plates, and $\Delta E(d)$ is the EMF for the distance $d$ in formulae 
(1-6). Naturally, if all the wings in the periodic configuration of plates 
are connected in parallel and not in series, it is necessary to calculate 
the equivalent summed current in the system.

\textbf{CALCULATION RESULTS }

From formula (\ref{eq7}) it follows that for $d=a$ in periodic configurations Casimir 
EMF is $\Delta E_{total} =\Delta E(a)$ at any $a$. It means that in the periodic structure, even for 
$d=a$, at $n\to \infty $ EMF is at the same level as that for one pair ( $n=1$) of the 
configuration. However, it is clear that at $d\ne a$, the periodic configuration 
EMF will depend on the $d/a$ ratio in accordance with formula (\ref{eq7}) for 
different $R/a$ ratios and angles $\varphi $ between the wings as it is shown in 
\hyperlink{fig3}{Fig.3a,c.} When the number $n$ of the pairs of wings grows, the behavior of the 
curve will be similar to those shown in \hyperlink{fig3}{Fig.3}, however, for any angles $\varphi$ and parameters $R/a$, however, naturally, the their EMF level will grow linearly depending on $n$. 

It is possible to determine the effectiveness $Q$ of the EMF generation in 
$n$ pairs of wings as the relation of $\Delta E_{total}$ to the total configuration length along the $z$-axis
\begin{equation}
\label{eq8}
Q=\frac{\Delta E_{total} }{n\left[ {a+d+2R\tan (\varphi )} \right]}.
\end{equation}
The dependence of $Q$ on the relations $R/a$ and $d/a$ is shown in \hyperlink{fig3}{Figs.3b.d.} It can be 
seen, for example, that for any length $R$ of the wings with different 
angles $\varphi$, there is the maximal effectiveness $Q$ of the generation of the 
total EMF $\Delta E_{total}$.

As it is known \cite{Fateev:2015}, there is the maximum of the EMF 
generation for each pair of wings depending on the angle $\varphi$ between the 
wings. The similar effect is also found for the periodic configuration \hyperlink{fig4}{Fig.4a.}
The combined dependences of EMF on two parameters, i.e. the angle $\varphi$ and the relation $d/a$, and also $R/a $ and $d/a$ are displayed in 
\hyperlink{fig5}{Fig.5a,c.}. The corresponding effectivenesses $Q$ are shown in \hyperlink{fig5}{Fig.5b,d}.

Let us note, that analytically and in the result of the numerical 
calculations, the value of the optimal relation $d/a\to 1.62$ at $\varphi \to 0$ and 
$R/a\to \infty $ has been found for the periodic configurations. At the 
combined search with the use of angles and relation $d/a$, the maximum of 
the EMF generation shifts to the values of the order $d/a\to 1.8$ at $\varphi 
=1\;\deg $.

\textbf{CONCLUSIONS}

Thus, in the present paper, the possibility in principle is shown for the 
existence of Casimir EMF in nanosized nonparallel metal wings arranged in 
periodic structures. It is found that in the periodic structures with pairs 
of nonparallel wings, all Casimir currents are compensated, and EMF is not 
generated. However, when the periodicity of the pairs of nonparallel wings 
is disturbed, uncompensated currents are generated in them in the direction 
of the smallest angle between the wings, and consequently, EMF is generated. 
In this case, at any relations of the configuration parameters (angles 
between the wings and the wings lengths, the distance between the wings, 
etc.) and at any number of the pairs of wings at their series connection in 
the chain, there is the maximum of the effectiveness of the EMF generation 
in the period. The value of the total EMF in the nonparallel pairs of wings 
connected in series linearly grows depending on the number of elements in 
the chain.
\begin{acknowledgments}
The author is grateful to T. Bakitskaya for hers helpful
participation in discussions.
\end{acknowledgments}

\end{document}